\documentclass[journal]{IEEEtran}

\usepackage{inconsolata}
\usepackage{graphicx}
\usepackage{amsmath}
\usepackage{amssymb}
\usepackage{booktabs}
\usepackage{url}
\usepackage{pifont}
\usepackage{xcolor}
\usepackage[caption=false]{subfig}
\usepackage{cleveref}

\newcommand{\etal}{\textit{et al.}}

\begin{document}

\title{Hear Me, See Me, Understand Me:\\ Audio-Visual Autism Behavior Recognition}

\author{
        Shijian~Deng,~\IEEEmembership{Member,~IEEE,}
        Erin E. Kosloski,
        Siddhi~Patel,
        Zeke A. Barnett,
        Yiyang~Nan, \\
        Alexander~Kaplan,
        Sisira~Aarukapalli,
        William T. Doan,
        Matthew~Wang, 
        Harsh~Singh, \\
        Pamela R. Rollins, Yapeng~Tian,~\IEEEmembership{Member,~IEEE} 

\thanks{S. Deng, Z. Barnett, A. Kaplan, S. Aarukapalli, W Doan, M. Wang, and Y. Tian are with the Department of Computer Science, The University of Texas at Dallas, Richardson, TX, 75080 USA.
}
\thanks{E. Kosloski and P. Rollins are with the School of Behavioral and Brain Sciences, The University of Texas at Dallas, Dallas, TX, 75235 USA 
.}
\thanks{Y. Nan is with the Department of Computer Science, Brown University, Providence, RI, 02912 USA 
.}
\thanks{H. Singh is with the Mohamed bin Zayed University of Artificial Intelligence, Abu Dhabi, UAE
.}
}

\maketitle

\begin{abstract}
In this article, we introduce a novel problem of audio-visual autism behavior recognition, which includes social behavior recognition, an essential aspect previously omitted in AI-assisted autism screening research. We define the task at hand as one that is audio-visual autism behavior recognition, which uses audio and visual cues, including any speech present in the audio, to recognize autism-related behaviors. To facilitate this new research direction, we collected an audio-visual autism spectrum dataset (AV-ASD), currently the largest video dataset for autism screening using a behavioral approach. It covers an extensive range of autism-associated behaviors, including those related to social communication and interaction. To pave the way for further research on this new problem, we intensively explored leveraging foundation models and multimodal large language models across different modalities. Our experiments on the AV-ASD dataset demonstrate that integrating audio, visual, and speech modalities significantly enhances the performance in autism behavior recognition. Additionally, we explored the use of a \textit{post-hoc to ad-hoc} pipeline in a multimodal large language model to investigate its potential to augment the model's explanatory capability during autism behavior recognition. We will release our dataset, code, and pre-trained models.
\end{abstract}

\begin{IEEEkeywords}
Audio, Video, Speech, Autism, Multimodal Large Language Model, Explainability, Dataset, Benchmark.
\end{IEEEkeywords}

\section{Introduction}

\IEEEPARstart{A}{utism} spectrum disorder (ASD) is a complex, heterogeneous neurodevelopmental condition associated with persistent challenges in social communication and interaction, as well as the presence of restrictive or repetitive behaviors and interests~\cite{dsm5tr}. The prevalence of autism has steadily risen in recent decades. Early identification and intervention are critical to support autistic\footnote{We use identity-first (\emph{i.e.,} autistic person) rather than person-first (\emph{i.e.,} person with autism) as preferred by most autistic self-advocates~\cite{bottema2021avoiding}. } children in developing social communication, language, and adaptive functioning~\cite{fuller2020effects, nahmias2019effectiveness}, yet the worldwide average age at which ASD is diagnosed is 60.48 months~\cite{van2021age}. Current early screening efforts primarily use parent report instruments (\textit{e.g.,} \cite{mchatrf2009}) which are suitable only for children under 30 months and are inherently subjective.

\begin{figure}[t]
  \centering
   \includegraphics[width=\linewidth]{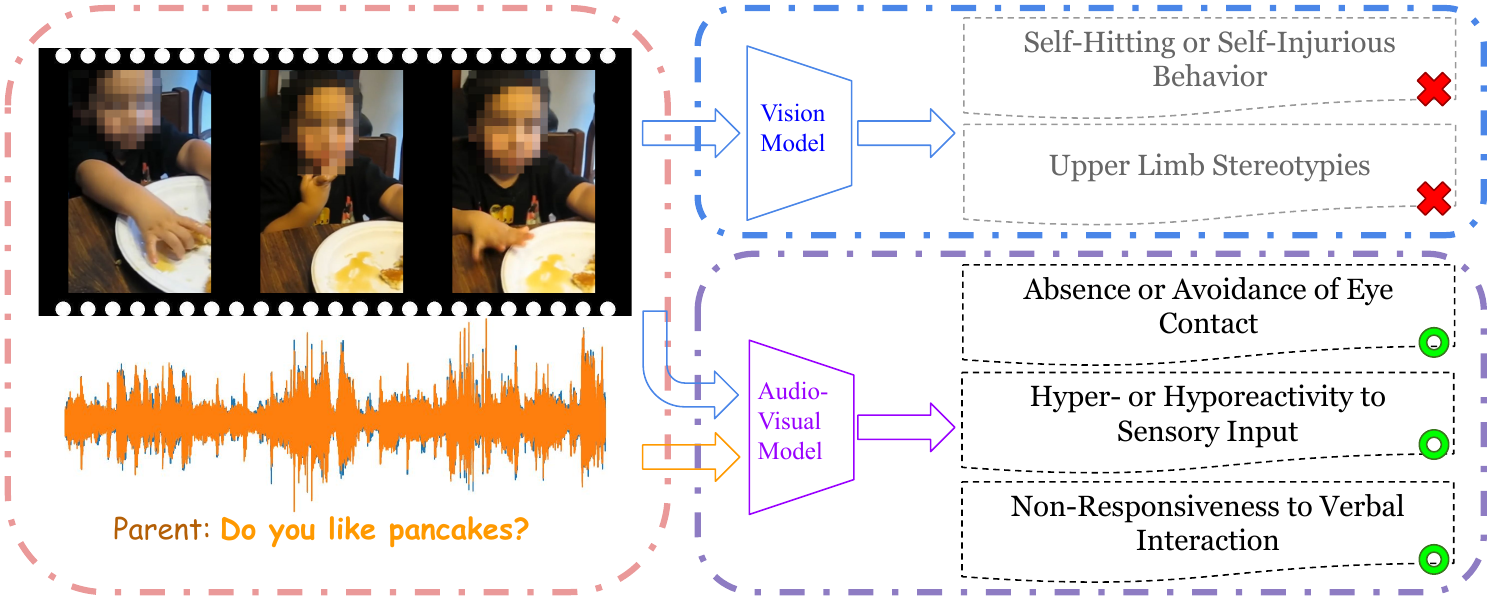}
   \vspace{-5mm}
   \caption{The vision-only model incorrectly identified two behaviors that were not present, whereas the audio-visual model correctly identified the three behaviors present in the clip. This illustrates how multimodal integration enables more accurate behavior identification.}
   \label{fig:teaser}
   \vspace{-5mm}
\end{figure}

Hence, there is a critical demand for an efficient, accessible, and objective screening tool that can effectively be used for a broader age range of children. To this end, researchers are exploring the application of artificial intelligence (AI) to autism screening. Previous studies have investigated the application of AI to autism detection by using various objective data-driven approaches such as functional Magnetic Resonance Imaging (fMRI) scans, eye tracking data, and behavioral observations  (\textit{e.g.,} \cite{supekar2022deep}, \cite{cilia2021computer}, \cite{megerian2022evaluation}). In our current study, we employ the behavioral observation method to screen autism, using the clinical gold standard criteria from the Diagnostic and Statistical Manual of Mental Disorders (DSM-V), \textit{i.e.}. This involves analyzing both social interaction challenges and restricted and repetitive behaviors (RRBs) from video datasets~\cite{dsm5tr}. Prior research that has utilized behavioral observation methods by using video datasets to identify autistic behaviors has focused solely on the visual modality. This singular focus has limited the scope of analyses as it tends to capture only RRB overlooking social interaction challenges integral to the diagnosis of autism. Further, RRBs are not always present before 36 months, whereas social challenges are detectable in the first year of life~\cite{jones2013attention, mundy2022bidirectional}. Detecting social interaction difficulties, however often requires moving beyond the visual modality and accessing auditory information. Thus it is imperative for an AI early autism screening tool to be capable of identifying behaviors related to autistic individuals' social interaction difficulties. Therefore, in an effort to better leverage AI for screening autism we created a new audio-visual autism behavior recognition dataset (AV-ASD) and introduced an audio-visual autism behavior recognition task, which aims to identify both social interaction behaviors in addition to RRBs. The AV-ASD dataset, featuring 928 video clips from 569 unique videos across 10 categories, is currently the largest for autism screening using a behavioral approach.

With the newly collected dataset, we establish a comprehensive benchmark for exploring how to better recognize autism behaviors in videos. We develop several baselines and novel frameworks using strong foundation models like CLIP (image), ImageBind (video/audio), and Whisper (speech). We further investigate the multimodal integration with temporal modeling and evaluate the effectiveness of Multimodal Large Language Models (MLLMs), including GPT-4V~\cite{openai2023gpt4} and LLaVA~\cite{li2023llava}, as zero-shot benchmarks. To utilize audio and speech cues in MLLMs, we adopt audio captioning and speech recognition models to generate text prompts. 
To further improve performance, we employ an audio-visual instruction tuning, adapting LLaVA into LLaVA-ASD with our annotated data. This significantly enhances its efficacy, particularly with audio-augmented prompts. However, solely relying on behavior labels during tuning can compromise the model's explainability and lead to catastrophic forgetting. To address these challenges, we propose a novel \textit{post-hoc to ad-hoc} framework that maintains the model's predictive accuracy while preserving its prediction explanation ability.

Our contributions are as follows: (1) We introduce a comprehensive autism behavior recognition dataset that contains 928 clips covering 10 categories annotated from 569 videos, which is the current largest dataset for autism screening in unconstrained videos. The dataset covers both restricted, repetitive patterns of behavior and unique social interaction behaviors relevant to autism, which were omitted in previous similar datasets;
(2) We extensively benchmark different state-of-the-art foundation models with diverse modality combinations and discover that specialized MLLMs by integrating audio-visual cues are capable of tackling autism behavior recognition. (3) Our \textit{post-hoc to ad-hoc} framework demonstrates potential in mitigating catastrophic forgetting during instruction tuning, while also achieving explainable predictions.

\begin{figure*}
  \centering
  \includegraphics[width=0.85\linewidth]{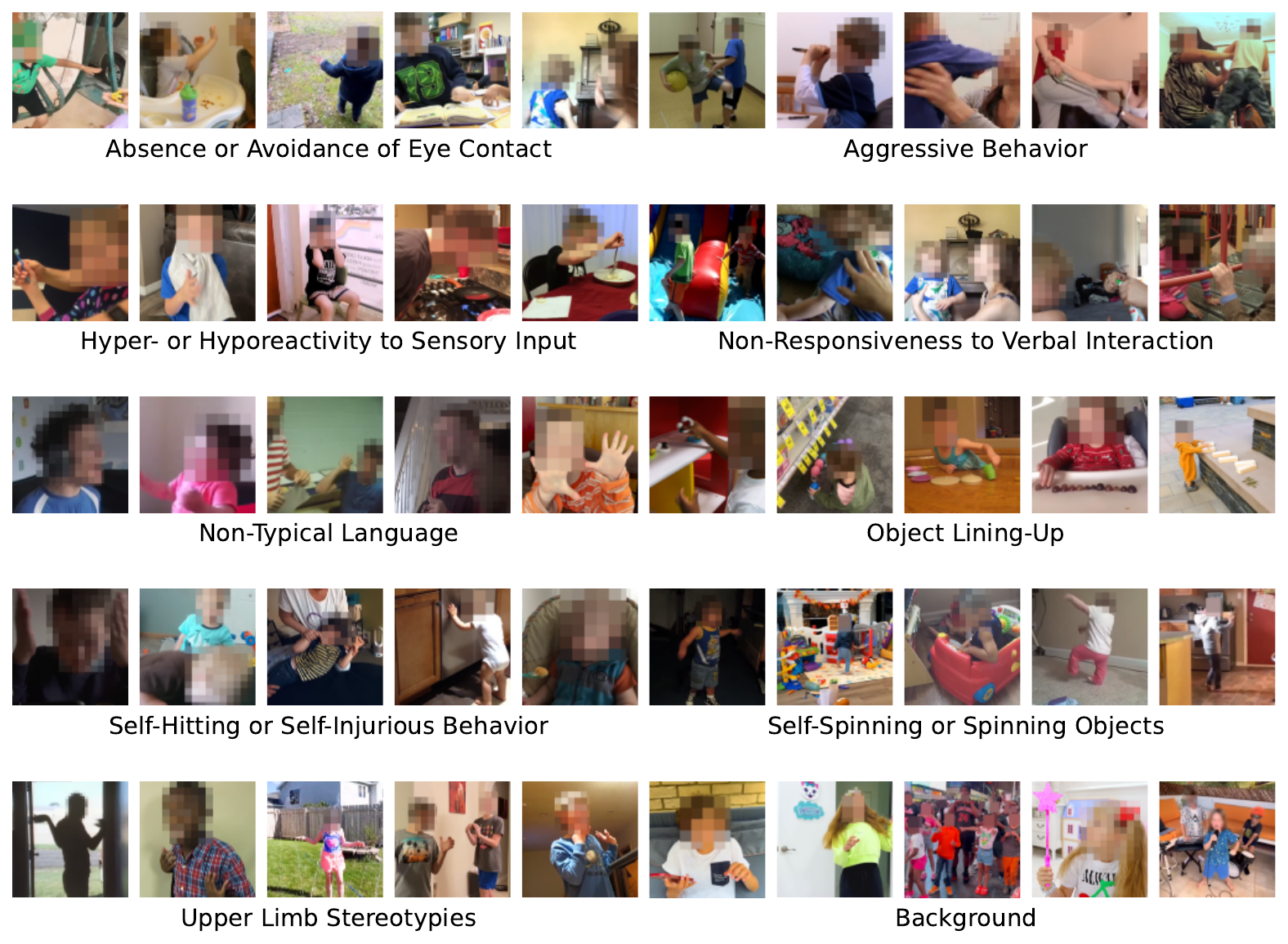}
  \vspace{-5mm}
  \caption{A depiction of the AV-ASD dataset, illustrating five sample instances from each category.}
  \label{fig:data_overview}
  \vspace{-5mm}
\end{figure*}

\section{Related Work}

\subsection{AI and Datasets in ASD Research}
Datasets play a crucial role in AI-powered autism detection. Traditionally, these datasets are comprised of videos recorded exclusively in a laboratory setting~\cite{billing2020dream, rehg2013decoding, riva2020enigma, zunino2018video, pandey2020guided, del2017study, dawson2018atypical, martin2018objective, li2023mmasd}. However, this approach inherently limits the datasets to controlled environments and scenarios.
There have been commendable strides towards introducing more realistic, unbounded datasets such as the Self-Stimulatory Behavior Dataset (SSBD)\cite{rajagopalan2013self}, the Expanded Stereotype Behavior Dataset (ESBD)\cite{negin2021vision}, the dataset curated by Wei \etal~\cite{wei2022vision}, and the Autism Stimming Behavior Dataset (ASBD)~\cite{ribeiro2023stimming}. Despite this progress, these datasets primarily focus on RRBs, neglecting the atypical social interactions necessary for autism diagnosis.  
Leveraging these self-stimulatory behavior and RRB datasets, researchers have proposed various vision-based models for early ASD detection~\cite{rajagopalan2013self, negin2021vision, wei2022vision, deng2022problem, ali2022video, rajagopalan2014detecting}. However, relying solely on vision overlooks the valuable social information that can be captured only with audio. Addressing these limitations, our work introduces a new multimodal ASD dataset and a novel multimodal learning framework to best leverage audio-visual cues.

\subsection{Audio-Visual Learning}
Audio-visual learning has shown its potential in various applications, including audio-visual action, emotion or speech recognition~\cite{kazakos2019epic, xiao2020audiovisual, gao2020listen, chen2022mm, lin2011error, tao2020end}, audio-visual localization~\cite{tian2018audio, xue2021audio, liu2022dense, huang2023egocentric, mo2023audio, jiang2023leveraging}, audio-visual video parsing~\cite{tian2020unified, mo2022multi}, audio-visual segmentation~\cite{zhou2022audio}, audio-visual source separation~\cite{zhao2018sound, gao2021visualvoice, su2023separating}, and audio-visual question
answering or dialogue~\cite{li2022learning, zhu2020describing, alamri2019audio}. The concurrent analysis of audio and visual data can provide a more holistic understanding of complex behaviors, particularly those in which the audio component plays a critical role. In this work, we have integrated audio data into our model, demonstrating that audio-visual learning can substantially enhance the efficiency of identifying a range of autistic behaviors. 

\begin{figure}[t]
  \centering
  \includegraphics[width=0.8\linewidth]{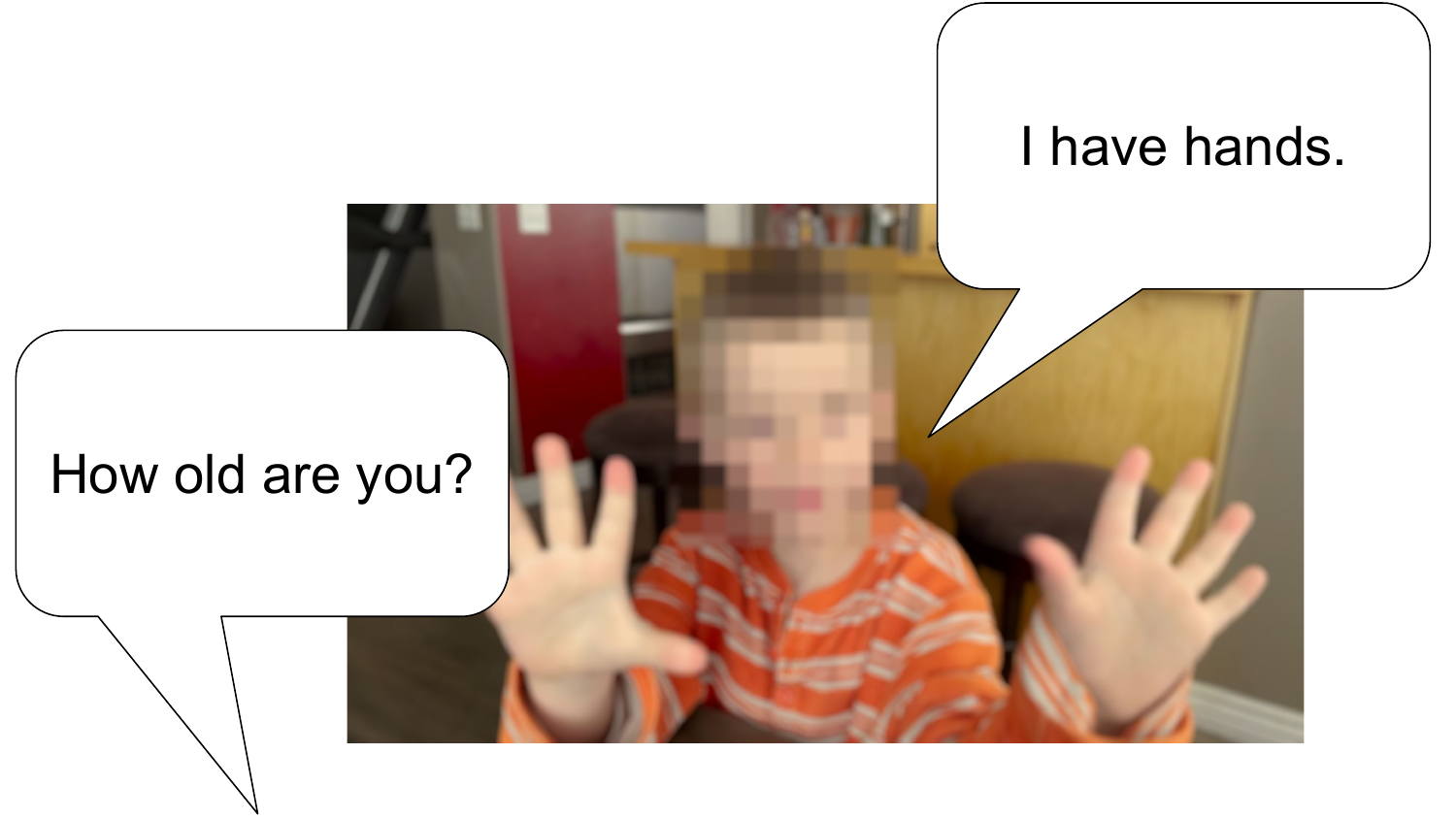}
  \vspace{-5mm}
  \caption{An ASD child responds ``I have hands.'' to the person who asks ``How old are you?''}
  \label{fig:hands}
  \vspace{-3mm}
\end{figure}

\begin{figure*}[t]
  \centering
  \subfloat[Category-wise distribution.]{\includegraphics[width=0.2\linewidth]{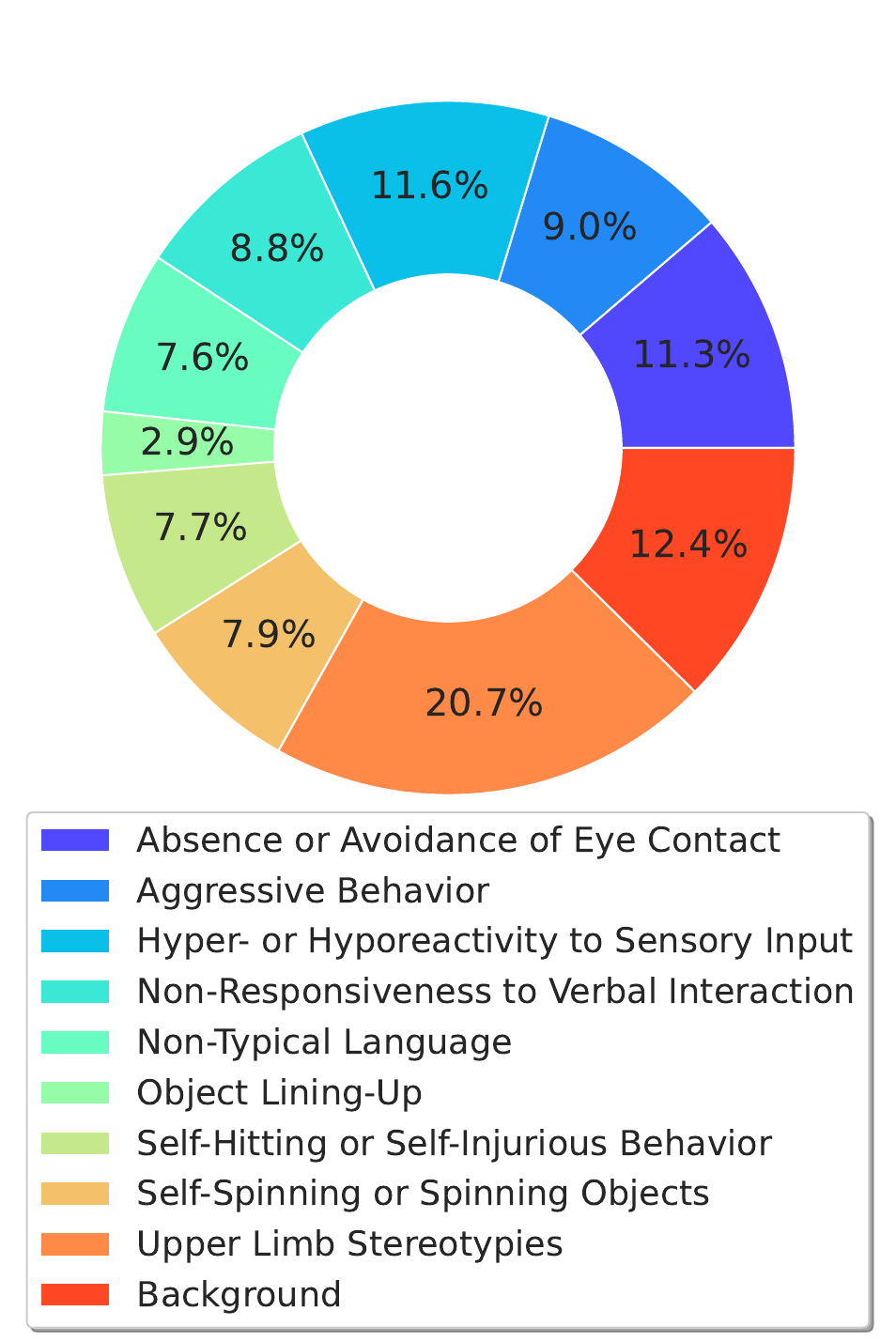}
    \label{fig:short-a}}
  \hfill
  \subfloat[Distribution of video in different splits.]{\includegraphics[width=0.75\linewidth]{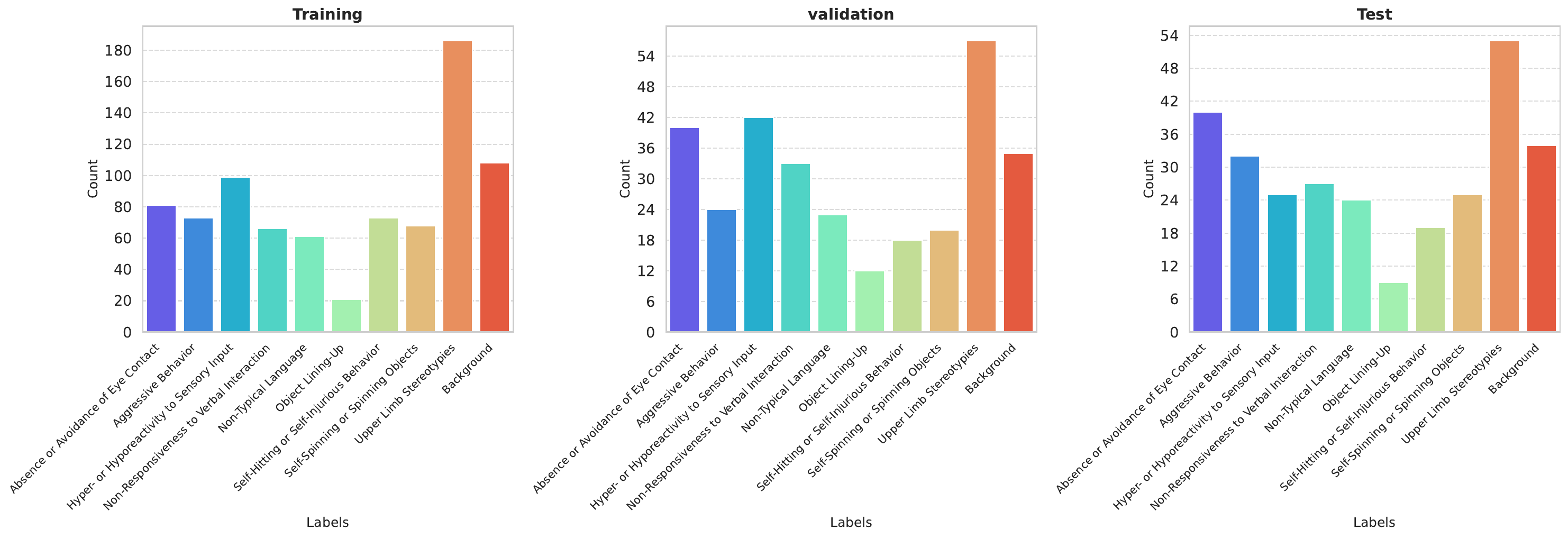}
    \label{fig:short-b}}
  \caption{Statistical illustrations of the AV-ASD dataset.}
  \label{fig:statistics}
\vspace{-5mm}
\end{figure*}

\section{The AV-ASD Dataset}

\begin{table}
  \centering
  \begin{tabular}{@{}l|cccc@{}}
    \toprule
    Dataset                             & Clips         & Categories  & Multi-Label                  & Social Behaviors  \\ \midrule
    SSBD~\cite{rajagopalan2013self}     & 75            & 3           & \textcolor{red}{\ding{55}}   & \textcolor{red}{\ding{55}} \\ 
    ESBD~\cite{negin2021vision}         & 141           & 4           & \textcolor{red}{\ding{55}}   & \textcolor{red}{\ding{55}} \\
    Wei \etal~\cite{wei2022vision}      & 61            & 3           & \textcolor{red}{\ding{55}}   & \textcolor{red}{\ding{55}} \\ 
    ASBD~\cite{ribeiro2023stimming}     & 165           & 4           & \textcolor{red}{\ding{55}}   & \textcolor{red}{\ding{55}} \\\midrule
    AV-ASD                              & \textbf{928}  & \textbf{10} & \textcolor{green}{\ding{51}} & \textcolor{green}{\ding{51}} \\ \bottomrule
  \end{tabular}
  \caption{Comparison of the AV-ASD dataset with other autism-related behavior datasets used in ASD screening research.}
  \vspace{-7mm}
  \label{tab:datasets}
\end{table}

The Audio-Visual Autism Spectrum Dataset (AV-ASD) is a curated collection designed to enhance research into autism-related behaviors, especially in social contexts. It includes 928 clips from 569 YouTube and Facebook videos, capturing diverse behaviors and environments. Some examples are shown in Fig.~\ref{fig:data_overview}.
AV-ASD distinguishes itself from preceding datasets in several significant ways, as delineated in Tab.~\ref{tab:datasets}. First, our dataset offers a far greater number of categories and video clips than all previous datasets combined. Second, AV-ASD is the first dataset to include social behaviors. Third, our dataset is the first ever autism dataset to use a multi-label setting, which is more practical since multiple autism-related behaviors could happen at the same time. Lastly, each instance in the AV-ASD dataset is meticulously annotated with time-stamped labels that identify various autistic behaviors. This stands in contrast to some previous datasets, such as ESBD, which are weakly labeled and lack specific start and end times for the observed behaviors. These make AV-ASD a pioneering dataset for AI-based multimodal autism research.

\subsection{Multimodal Nature of Social Behaviors}
\label{social}

Detection and interpretation of social behaviors often require both auditory and visual information. In particular, the context of social interactions is deeply intertwined with the content of conversations. As such, audio, specifically speech, offers an essential layer of contextual data invaluable for social behavior analysis. For example, as shown in Fig.~\ref{fig:hands}, a child answering another person's question could be visually interpreted as a normal communicative engagement. Yet, when the speech content is analyzed, it is clear that the child's response bears no relevance to the posed question, illustrating the critical role of multimodal analysis.

Thus, our multimodal framework aids in constructing a more accurate model of social autistic behaviors.

\subsection{Data Collection, Annotation, and Statistics}
To create our behavioral categories for the social and RRB domains of autism, we identified social challenges from the social behavior classifications of the DSM-V-TR~\cite{dsm5tr} and M-CHAT-R/F~\cite{mchatrf2009} screening tool. RRB behaviors were adapted from SSBD~\cite{rajagopalan2013self} and ESBD~\cite{negin2021vision}. Notably, we were limited to behaviors that could be identified in a brief video clip. The resulting taxonomy consisted of nine distinct autistic behavioral categories and one \textit{Background} (i.e., not-applicable) category. 

We curated the AV-ASD dataset through a keyword video search and excluded irrelevant content such as lectures and cartoons. This resulted in 928 distinct video clips extracted from 569 online videos. A team of six volunteer students meticulously annotated each clip, followed by verification by a Speech Pathologist (SLP) with 15 years of experience working with autistic children.

Our final AV-ASD dataset consists of 928 clips in 10 categories, amounting to roughly 6 hours and 40 minutes of footage, extracted from 569 videos with a total duration exceeding 86 hours. A detailed statistical analysis of our dataset is provided in Fig.~\ref{fig:statistics}. For each category: 
\textit{Absence or Avoidance of Eye Contact} includes 161 clips;
\textit{Aggressive Behavior} comprises 129 clips;
\textit{Hyper- or Hyporeactivity to Sensory Input} incorporates 166 clips;
\textit{Non-Responsiveness to Verbal Interaction} contains 126 clips;
\textit{Non-Typical Language} has 108 clips.
\textit{Object Lining-Up} incorporates 42 clips;
\textit{Self-Hitting or Self-Injurious Behavior} comprises 110 clips;
\textit{Self-Spinning or Spinning Objects} includes 113 clips;
\textit{Upper Limb Stereotypies} consists of 296 clips; and
We also added a \textit{Background} category that contains 177 clips. All clips were thoroughly reviewed for their relevance and validity in relation to our study. We randomly divided the dataset into a training set (553 clips), a validation set (193 clips), and a testing set (182 clips). Clips have a mean duration of 25.88s, median duration of 10.00s, maximum duration of 887.01s, and minimum duration of 1.00s. Each clip has an average of 1.54 behavior categories.

\subsection{Annotators and Instructions for Annotation}
All student annotators are residents of either the United States, China, or India. Out of the 6 students, 5 are male: 3 are from the United States, including one female, 2 from China, and 1 from India. The SLP is a female from the United States.

To acquire annotations, we implemented a two-step process:
\begin{enumerate}
    \item We instructed student annotators to review a given video and check for any signs of autistic behavior. If present, they were to label it in a single-label setting with the most significant category and record the start and end times of the behavior.
    \item We extracted clips from the video for fine-grained labeling. The SLP's task was to re-label the clips in a multi-label setting, confirming whether each autistic behavior appeared in each clip, without the need to record the start and end times again.
\end{enumerate}

\section{Autism Behavior Recognition}
\label{methods}

Building on recent advances in large models, we develop a novel set of methods exploiting both audio and visual modalities to identify autism-related behaviors in videos. These models are benchmarked on the AV-ASD dataset to evaluate their performance and establish baselines for future research.

We leverage recent foundation models to extract features for autism behavior recognition, utilizing CLIP~\cite{radford2021learning} for image features, ImageBind~\cite{girdhar2023imagebind} for processing images, videos, and audio, and Whisper~\cite{radford2023robust} for speech analysis. These extracted features are then inputted into either linear probes, Multi-Layer Perceptrons (MLPs), or temporal models for prediction.

\vspace{2mm}
\noindent
\textbf{Image Representation.}
To utilize CLIP and ImageBind as image encoders, a video is transformed into a single composite image. Specifically, given a sequence of frames \( V = \{F_1, F_2, \ldots, F_n\} \), nine frames are uniformly selected and arranged into a \(3\times3\) grid to form a composite image denoted as \(I_{V}\). CLIP or ImageBind then extracts features from this image, represented by \(f_{\text{CLIP}}(I_{V})\) for CLIP and \(f_{\text{ImageBind}}(I_{V})\) for ImageBind, respectively. These extracted features are subsequently utilized in logistic regression or MLP models designed to classify various autism behaviors. Additionally, this composite image format serves as instrumental visual input for MLLMs.

\vspace{2mm}
\noindent
\textbf{Audio and Video Representation.}
For both audio and video, we employ clip-level and segment-level representations. In the segment-level approach, an input video with accompanying audio is first divided into \(T\) non-overlapping pairs of visual and audio segments, denoted as \( \{V_t, A_t\}_{t=1}^{T} \), with each segment lasting 1 second. During the training phase, each segment is associated with a behavioral label \(y_t\). The audio feature vector for each audio segment \(A_t\) is encoded by \(f_{a_t} = \text{ImageBind}(A_t)\), and similarly, the video feature vector for each video segment \(V_t\) is encoded by \(f_{v_t} = \text{ImageBind}(V_t)\), utilizing the pre-trained ImageBind model.

\vspace{2mm}
\noindent
\textbf{Speech Representation.} To encode each audio segment \(A_t\) into a speech feature vector \(f_{s_t} = \text{pool}(\text{Whisper}(A_t))\), we utilize the encoder of the Whisper model~\cite{radford2023robust}, applying an average pooling over the time dimension at the end.

\vspace{2mm}
\noindent
\textbf{Temporal Modeling.}
For each modality \(m\) (e.g., audio, visual, speech), we process the segment-wise features for temporal modeling. These features are aggregated over adjacent ten segments to form feature sequences, represented as \(T_{t \to t+9}^m = T(F_{t \to t+9})\). Subsequently, a transformer encoder layer is applied to the sequence of each modality. This approach effectively integrates temporal information, resulting in a temporally aggregated feature representation denoted by \(x_m^{T}\). Along with features extracted directly from the entire clip without temporal modeling, all features are collectively represented as \(x_m\).

\vspace{2mm}
\noindent
\textbf{Multimodal Fusion.}
To integrate information across modalities, we leverage several multimodal fusion methods: average, max, concatenation, and weighted fusion.

\begin{itemize}

\item \textbf{Average Fusion:} The fused feature vector \(f\) is computed as the average of features \(x_m\) across all modalities \(m\). This is denoted as:
    \begin{equation}
    f = \frac{1}{M}\sum_{m=1}^{M}x_m,
    \end{equation}
where \(M\) represents the total number of modalities.

\item \textbf{Max Fusion:} The fused feature vector \(f\) is computed as the maximum feature \(x_m\) across all modalities \(m\):
    \begin{equation}
    f = \max_{m=1}^{M}x_m.
    \end{equation}

\item \textbf{Concatenation Fusion:} The fused feature vector \(f\) is computed by concatenating features \(x_m\) from all modalities \(m\). This is denoted as:
    \begin{equation}
    f = [x_1, x_2, \dots, x_M].
    \end{equation}

\item \textbf{Weighted Fusion:} The fused feature vector \(f\) is computed as the weighted average of features \(x_m\) across all modalities \(m\) with weights \(w_m\). This is denoted as:
\begin{equation}
f = \sum_{m=1}^{M}w_m \cdot x_m,
\end{equation}
where \(M\) represents the total number of modalities.

\end{itemize}

\begin{figure}[t]
  \centering
  \includegraphics[width=\linewidth]{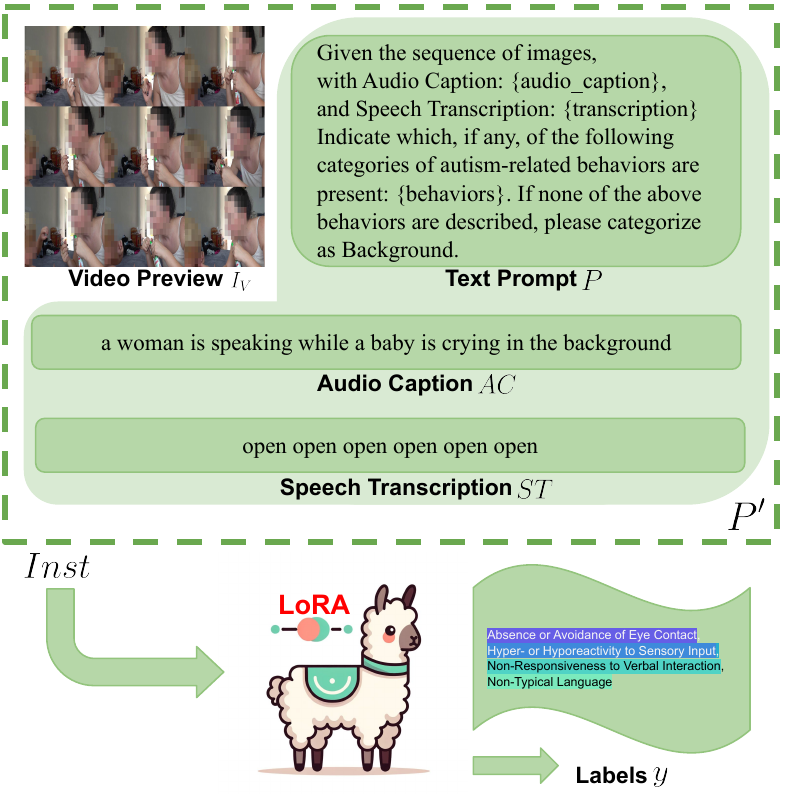}
  \vspace{-7mm}
  \caption{LLaVA-ASD: Instruction Tuning for LLaVA. Given a video preview $I_V$ and an enhanced text prompt $P'$, which is a text prompt $P$ augmented with an audio caption $AC$ and speech transcription $ST$. These elements are combined to form the model's instruction input $Inst$. The output consists of multiple autism behavior labels presented in text format as $y$. We employed LoRA for efficient fine-tuning.}
  \label{fig:lora}
  \vspace{-5mm}
\end{figure}

\subsection{Zero-shot Baselines with MLLMs }
MLLMs have revolutionized zero-shot learning, seamlessly integrating information across modalities like vision and language. To this end, we investigated the potential of MLLMs in precisely identifying autism behaviors in videos, focusing on their zero-shot capabilities. We employed two MLLMs: GPT-4V, a state-of-the-art proprietary model developed by OpenAI~\cite{openai2023gpt4}, and LLaVA, an open-source alternative excelling in similar tasks~\cite{liu2023llava, liu2023improvedllava} as the benchmark. 

\vspace{2mm}
\noindent
\textbf{MLLMs for Autism Behavior Recognition.} Given that current open-source SOTA MLLMs lack the ability to process long video sequences, we opted to repurpose the composite image, \(I_V\), as the visual input for our MLLMs.
 To assist the MLLMs in accurately identifying autism-related behaviors, we developed a textual prompt \( P \) (see Fig.~\ref{fig:lora}). This prompt is strategically devised to act as a linguistic guide, steering the MLLMs toward recognizing autistic cues and patterns inherent in the video content. The prediction of behavior is thus derived using the formula \( \hat{y} = \text{MLLM}(I_{V}, P) \), where \( \hat{y} \) represents the MLLMs' output.

\vspace{2mm}
\noindent
\textbf{Bridging the Multimodal Gap.}  
Current MLLMs typically focus on image and text inputs, posing a challenge for analyzing multimodal data including audio and speech cues. To overcome this limitation in our autism behavior recognition task, we propose a Multimodal Representational Text Fusion with two key strategies: (1) Audio Captioning~\cite{labbe2023conette}: It transformed audio segments into textual descriptions, enriching the input with semantic information extracted from the audio content; (2) Speech Recognition~\cite{radford2023robust}: This approach transcribed spoken segments into text, providing the model with direct linguistic cues from the audio modality.
By leveraging these strategies, we translate audio information into structured text representations readily usable by MLLMs. Combining this textual data with the prompt $P$ enables comprehensive multimodal analysis, empowering MLLMs to capture multimodal cues in video content, ultimately leading to a more accurate recognition of autism-related behaviors.

\vspace{2mm}
\subsection{MLLMs with Instruction Tuning}

Beyond zero-shot testing, we leverage the power of instruction tuning~\cite{ouyang2022training, dai2023instructblip, zhang2023video, liu2023visual} to enhance MLLMs' performance on our specialized dataset of video and audio data containing autism-related behaviors. This aims to refine the models' understanding of autism-specific cues from different modalities, leading to heightened effectiveness in identifying autism-related behaviors. Since GPT-4V is not open-source, we adopt LLaVA as a baseline for the instruction tuning study.

For training, we construct an instruction tuning pair denoted as \( \{Inst, y\} \). Here, \( Inst = [I_{V}, P'] \) represents the model's input for a video clip, where \( P' \) is an enhanced text prompt. This prompt \( P' \) is a combination of the initial input prompt \( P \) augmented with audio caption (AC) and speech transcription (ST). The term \( y \) refers to the annotated label for the behavior categories in the video clip. In implementation, we employed the low-rank adaptation (LoRA)~\cite{hu2021lora} for efficient training. We denote the trained model as LLaVA-ASD.

\section{Experiments}

\subsection{Experimental Setup}

\begin{table}
  \centering
  \footnotesize
    \begin{tabular}{l|c}
    \toprule
        Method                                      & F1-score (\%)\\ 
        \midrule
        Dummy Baseline                             & 26.83 \\ 
        \midrule
        CLIP  (ViT-L/14@336px, linear probe)        & 45.03 \\
        CLIP  (ViT-L/14@336px, MLP)                 & 45.72 \\
        \midrule
        ImageBind (image, linear probe)             & 12.12\\
        ImageBind (image, MLP)                      & 39.19 \\
        \midrule
        ImageBind (video, linear probe)             & 10.14 \\
        ImageBind (video, MLP)                      & 44.06 \\
        ImageBind (video, temporal)                 & 50.87 \\
        \midrule
        ImageBind (audio, linear probe)             & 27.54 \\
        ImageBind (audio, MLP)                      & 28.47 \\ 
        ImageBind (audio, temporal)                 & 41.74 \\
        \midrule
        Whisper (speech, linear probe)              & 34.04 \\
        Whisper (speech, MLP)                       & 36.48 \\
        Whisper (speech, temporal)                  & 39.69 \\
        \midrule
        GPT-4V zero-shot (vision)                   & 16.49 \\ 
        GPT-4V zero-shot (vision, audio)            & 19.52 \\
        GPT-4V zero-shot (vision, speech)           & 33.88 \\
        GPT-4V zero-shot (vision, audio, speech)    & 28.93 \\
        LLaVA zero-shot (vision)                    & 13.26 \\
        LLaVA zero-shot (vision, audio)             & 4.60 \\
        LLaVA zero-shot (vision, speech)            & 15.61 \\
        LLaVA zero-shot (vision, audio, speech)     & 11.61 \\
        \midrule
        ImageBind + Whisper (linear probe)          & 37.36 \\
        ImageBind + Whisper (MLP)                   & 39.97 \\
        ImageBind + Whisper (temporal)              & \underline{53.17} \\
        Ours (LLaVA-ASD)                            & \textbf{59.77} \\
    \bottomrule
    \end{tabular}
  \caption{Autism behavior recognition results of different baselines on AV-ASD test set. The \textbf{best} and \underline{second best} results are highlighted.}
  \vspace{-5mm}
  \label{tab:experiments_all}
\end{table}

\begin{table*}[t]
  \centering
  \footnotesize
    \begin{tabular}{@{}l|ccc|ccc@{}}
    \toprule
        Method                 & Average                &  Max      & Concat  &  $(a:v:s = 2:1:1)$  & $(a:v:s = 1:2:1)$ & $(a:v:s = 1:1:2)$ \\ 
        \midrule
        Audio                  &       41.74            &  N/A      &  N/A    &   N/A               & N/A               & N/A\\ 
        Visual                 &       50.87            &  N/A      &  N/A    &   N/A               & N/A               & N/A\\ 
        Speech                 &       39.69            &  N/A      &  N/A    &   N/A               & N/A               & N/A\\ 
        \midrule
        Audio-Visual           &       51.46            &  52.06    &  52.22  &   N/A               & N/A               & N/A\\ 
        Audio-Speech           &       42.13            &  42.80    &  42.28  &   N/A               & N/A               & N/A\\ 
        Visual-Speech          &       51.63            &  51.40    &  52.22  &   N/A               & N/A               & N/A\\ 
        \midrule
        Audio-Visual-Speech    &     \underline{52.70}  &  51.76    &  51.96  &   50.83             & 52.56             & \textbf{53.17}\\ 
    \bottomrule
    \end{tabular}
  \caption{F1-score (\%) on AV-ASD autism-related behaviors recognition results with different modalities and fusion methods (for single modality, no fusion is required, but we provide the performance in the average column for convenience of comparison). The last three columns represent weighted ratio sets that we use for the weighted fusion. The \textbf{best} and \underline{second best} results are highlighted.}
  \vspace{-5mm}
  \label{tab:experiments_temporal_modling}
\end{table*}

\begin{table*}[t]
  \centering
  \scalebox{1.}{
  \begin{tabular}{l|c|c|c|c}
  \toprule
  \textbf{Behavior}                         & \textbf{V} & \textbf{V+A} & \textbf{V+S} & \textbf{V+A+S } \\
  \midrule
  Absence or avoidance of eye contact       & 46.15 & 47.89 & \textbf{56.34} & \underline{55.38} \\
  Aggressive behavior                       & \underline{72.13} & 63.33 & 66.67 & \textbf{75.00} \\
  Hyper- or hyporeactivity to sensory input & \textbf{40.68} & 31.17 & \underline{35.14} & 29.03 \\
  Non-responsiveness to verbal interaction  & 36.36 & 33.96 & \textbf{48.28} & \underline{40.00} \\
  Non-typical language                      & 20.69 & 29.27 & \textbf{45.45} & \underline{32.43} \\
  Object lining-up                          & 75.00 & 82.35 & \underline{85.71} & \textbf{88.89} \\
  Self-hitting or self-injurious behavior   & \underline{50.00} & 40.00 & \textbf{52.63} & 43.90 \\
  Self-spinning or spinning objects         & 56.60 & 57.69 & \underline{60.38} & \textbf{65.38} \\
  Upper limb stereotypies                   & 57.45 & 58.06 & \underline{66.02} & \textbf{67.33} \\
  Background                                & 79.45 & \textbf{81.69} & \underline{81.08} & 81.01 \\
  \midrule
  \textbf{Average}                          & 53.45 & 52.54 & \textbf{59.77} & \underline{57.84} \\
  \bottomrule
  \end{tabular}
  }
  \caption{Autism behavior recognition results with different modalities by LoRA fine-tuned on LLaVA. The \textbf{best} and \underline{second best} results are highlighted.}
  \label{tab:experiments_categories_llava}
\end{table*}

For benchmarking purposes, we conducted a performance evaluation of several baseline models using our newly collected AV-ASD dataset. Below are the experimental details for these baselines.

\vspace{2mm}
\noindent
\textbf{Features Extraction.}
For image features, we utilized CLIP's top-performing model, ViT-L/14@336px, for visual feature extraction. For audio and video, the ImageBind huge model~\cite{girdhar2023imagebind} was employed to extract a 1024-dimensional (1024-D) feature from each clip. Additionally, we used ImageBind to extract image features as well, facilitating a comparison with CLIP. For speech, Whisper's medium model~\cite{radford2023robust} was used, producing a 1024-D embedding from the encoder after pooling, thus matching the feature size extracted from ImageBind. This feature size compatibility enables effective feature fusion in later stages.

We used the CoNeTTE~\cite{labbe2023conette}, pretrained on Clotho~\cite{drossos2020clotho}, to generate audio captions (AC) and employed Whisper large-v3~\cite{radford2023robust} to generate speech transcriptions (ST) for each clip.

\vspace{2mm}
\noindent
\textbf{Linear Probes and Non-Linear Mapping.}
A logistic regression model was trained with the settings: random state set to 0 and maximum iterations limited to 3000, aiming to predict behavior categories based on each modality feature individually. We further explored multimodal fusion through four methods: average, max, concatenation, and weighted fusion. For the latter, we allowed the weights for each modality to be either 1.0 or 2.0, before applying linear probes to the fused features.
Transitioning from logistic regression, we employed an MLP for nonlinear mapping using the MLPClassifier from scikit-learn, with the random state set to 0 and maximum iterations at 3000. The same fusion strategies were applied, and the outcomes were documented in \cref{tab:experiments_all}.

\begin{figure}[t]
  \centering
  \includegraphics[width=\linewidth]{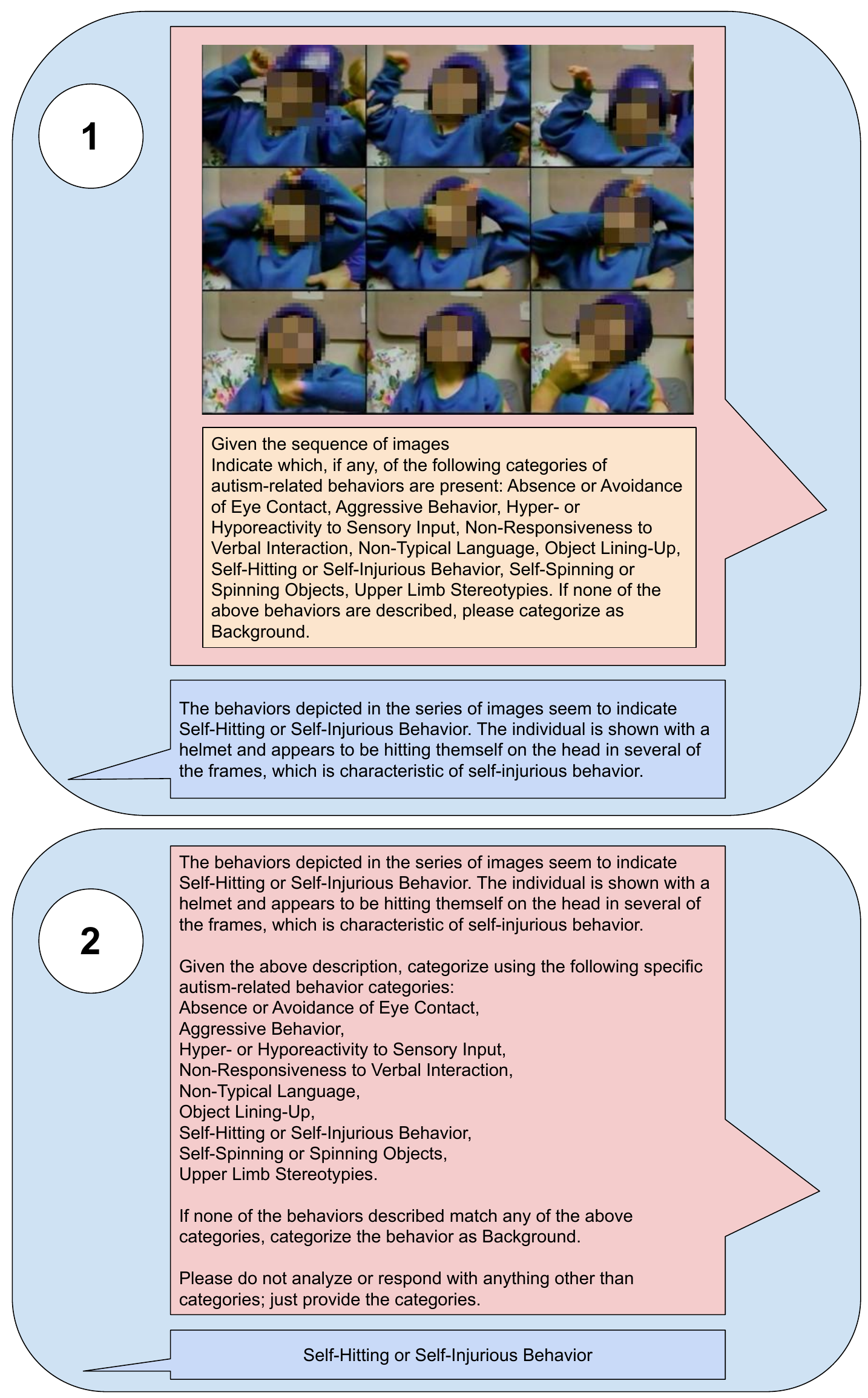}
  \vspace{-7mm}
  \caption{The prompt we use to obtain zero-shot inference results from GPT-4V and LLaVA. Two steps are separately presented in two boxes. The red messages are user inputs and the blue messages are model responses.}
  \label{fig:prompt_zero_shot}
  \vspace{-5mm}
\end{figure}

\vspace{2mm}
\noindent
\textbf{Temporal Modeling.}
Using ImageBind, we extracted a 1024-dimensional feature from a one-second audio segment and another 1024-dimensional feature from a one-second visual segment. Additionally, Whisper’s medium model was deployed to extract a 1024-dimensional speech feature from a one-second audio segment.
For temporal modeling, we focused on a time sequence of 10 seconds, concatenating features from each second to form a sequence spanning 10 seconds in length. An intensive hyperparameter search was conducted, experimenting with various numbers of nodes, layers, and transformer layer heads, as well as dropout rates, weight decay factors, etc. We utilized the Adam optimizer with an initial learning rate of 0.001 and trained all models for 100 epochs with a batch size of 1024. Cross-entropy loss with class weights was employed to tackle imbalance issues.

\vspace{2mm}
\noindent
\textbf{MLLMs.}
To test the zero-shot performance of MLLMs, we used the OpenAI API to get the response of GPT-4V with the specific checkpoint of gpt-4-vision-preview\footnote{https://platform.openai.com/docs/guides/vision}. For LLaVA, we used the Hugging Face checkpoint llava-v1.5-13b\footnote{https://huggingface.co/liuhaotian/llava-v1.5-13b}.

The prompt we used to obtain inference results from GPT-4V and LLaVA is shown in Figure~\ref{fig:prompt_zero_shot} and involves a two-step pipeline:
\begin{enumerate}
    \item We utilized an image and the task description to request the model to recognize autism behaviors. However, oftentimes, the model adopted a freestyle approach and did not directly provide behavior labels.
    \item We used the output from the previous model as input and sent it to GPT-4V, allowing it to reformat the answer into a simple behavior label format. This enabled us to easily evaluate performance through word matching.
\end{enumerate}

\begin{figure*}[t]
  \centering
  \includegraphics[width=\linewidth]{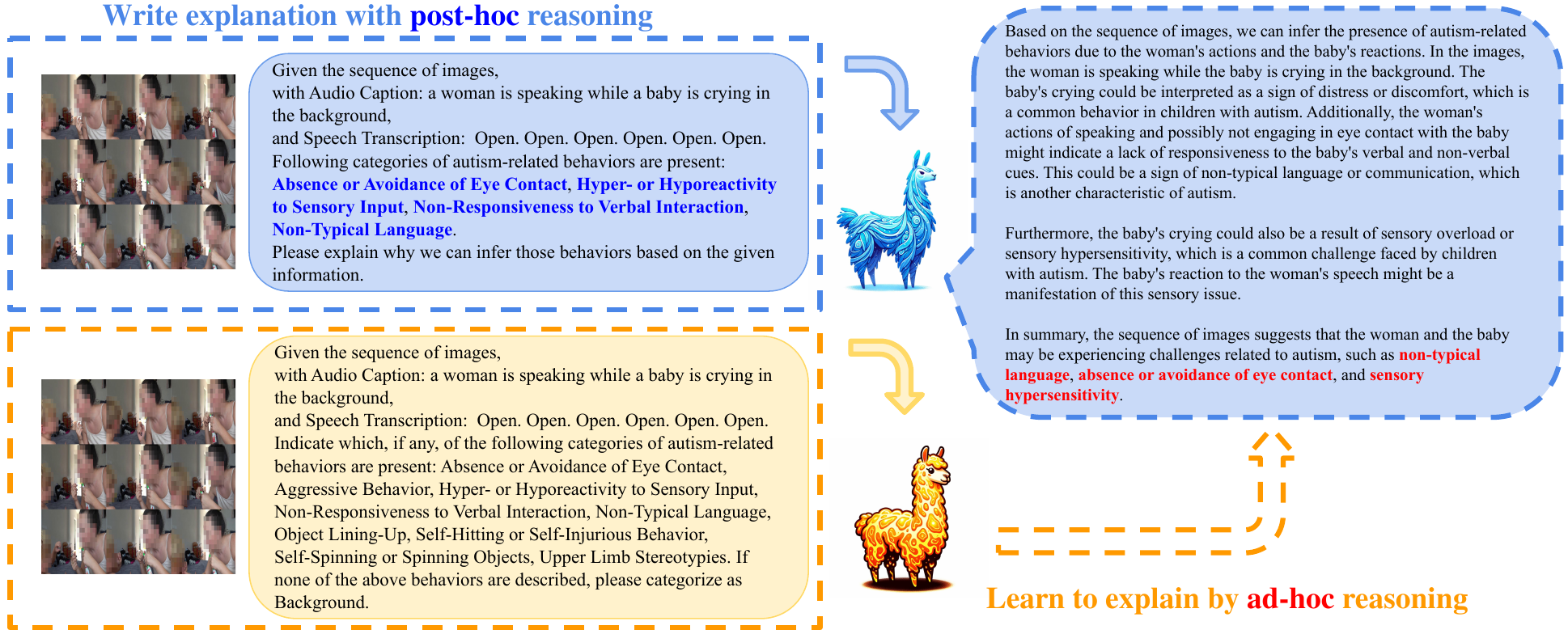}
  \vspace{-5mm}
  \caption{Explainability Framework of \textit{post-hoc to ad-hoc}. The model initially generates a pseudo-explanation (\textit{Post-Hoc}) based on the provided ground truth. Subsequently, it uses this pseudo-explanation as guidance to learn how to explain the decision-making process in identifying behaviors without the need for provided ground truth (\textit{Ad-Hoc}).}
  \vspace{-5mm}
  \label{fig:exp}
\end{figure*}

For fine-tuning the LLaVA model, we enabled LoRA with $r = 128$ and $\alpha = 256$. The learning rate for the multimodal projector was set at $2 \times 10^{-5}$. The model was trained using DeepSpeed with configurations from \texttt{zero3}. CLIP ViT-L/14@336px was employed as the vision encoder. Each model underwent training for 100 epochs, with inference conducted on the validation dataset every 10 epochs to identify the optimal epoch. The models’ performance on the testing dataset is documented in Table~\ref{tab:experiments_categories_llava}.

\vspace{2mm}
\noindent
\textbf{Evaluation Metric.}
We used the macro-averaged F1 score to evaluate model performance. For a fair comparison, all models were trained on the same training set of the AV-ASD dataset.

\subsection{Results and Analysis}
Table~\ref{tab:experiments_all} compares the recognition performance of various baselines.
Table~\ref{tab:experiments_temporal_modling} provides the ImageBind + Whisper model's performance with different fusion strategies.
Table~\ref{tab:experiments_categories_llava} provides LLaVA results that show the impact of different modalities.

\vspace{2mm}
\noindent
\textbf{Vision Matters.} The CLIP model achieves an F1 score exceeding 45\%, significantly outperforming the dummy baseline, which assigns a positive prediction $1$ to all predictions. This indicates that autism behaviors exhibit distinct visual patterns recognizable by vision perception models. Additionally, Table~\ref{tab:experiments_all} shows that ImageBind (video, temporal modeling) gains higher accuracy compared to temporal models utilizing audio and speech features alone. These findings underscore the critical role of visual features in autism behavior recognition.

\vspace{2mm}
\noindent
\textbf{Multimodal integration is helpful.} From Table~\ref{tab:experiments_all}, we observe that the multimodal model, ImageBind + Whisper, surpasses the performance of unimodal models using solely ImageBind or Whisper features. These results highlight the benefits of integrating cues from audio, visual, and speech modalities in recognizing autism behaviors in videos. Additionally, the findings in Table~\ref{tab:experiments_temporal_modling} and Table~\ref{tab:experiments_categories_llava} further substantiate the efficacy of multimodal fusion. Although integrating ambient audio can sometimes be beneficial, we notice it is not a straightforward `plug and play' solution. For example, when we use audio captions in LLaVA, they do not enhance the overall performance of the task. This suggests that the current audio captions may not be sufficient, and a more effective representation of ambient audio may need to be explored in future work to improve MM-LLM performance in detecting autism-related behaviors.

\vspace{2mm}
\noindent
\textbf{Zero-shot testing fails.} 
Initially, we tested the zero-shot ability of GPT-4V and LLaVA to identify autistic behavior, but both fell short as in Table~\ref{tab:experiments_all}. GPT-4V simply refused to answer most of the requests, and LLaVA predominantly predicted \textit{Background}. The results demonstrate the two MLLMs cannot directly be used to tackle our task.

\vspace{2mm}
\noindent
\textbf{Instruction tuning with LLaVA works.} 
We conducted further fine-tuning of LLaVA under four settings: vision only, audio-visual, visual-speech, and audio-visual-speech, to evaluate how different modalities contribute to the recognition performance (see Table~\ref{tab:experiments_categories_llava}). (1) A noteworthy observation is that although LLaVA's visual encoder is identical to CLIP's, the V-only model's performance (53.45\%) significantly surpasses that of the original CLIP equipped with a linear probe (45.03\%). This improvement indicates that the LLM component in LLaVA effectively boosts the perception encoder's classification efficacy. The fine-tuned model outperformed all previous baselines listed in Table~\ref{tab:experiments_all}, thereby illustrating the superior capability of MLLMs in identifying autism behaviors. (2) The LLaVA-ASD model, incorporating visual, audio, and speech information, achieves the best performance in many categories, while combining visual and speech data together achieved the best overall performance, emphasizing the importance of multimodal data for accurate recognition of autism behaviors. This suggests that the use of audio captioning and speech transcription can leverage auditory and speech information into prompts, thus improving recognition in videos. Additionally, these results confirm that the MLLM serves as an efficient integrator of different modalities.

\begin{figure}[t]
  \centering
  \includegraphics[width=1\linewidth]{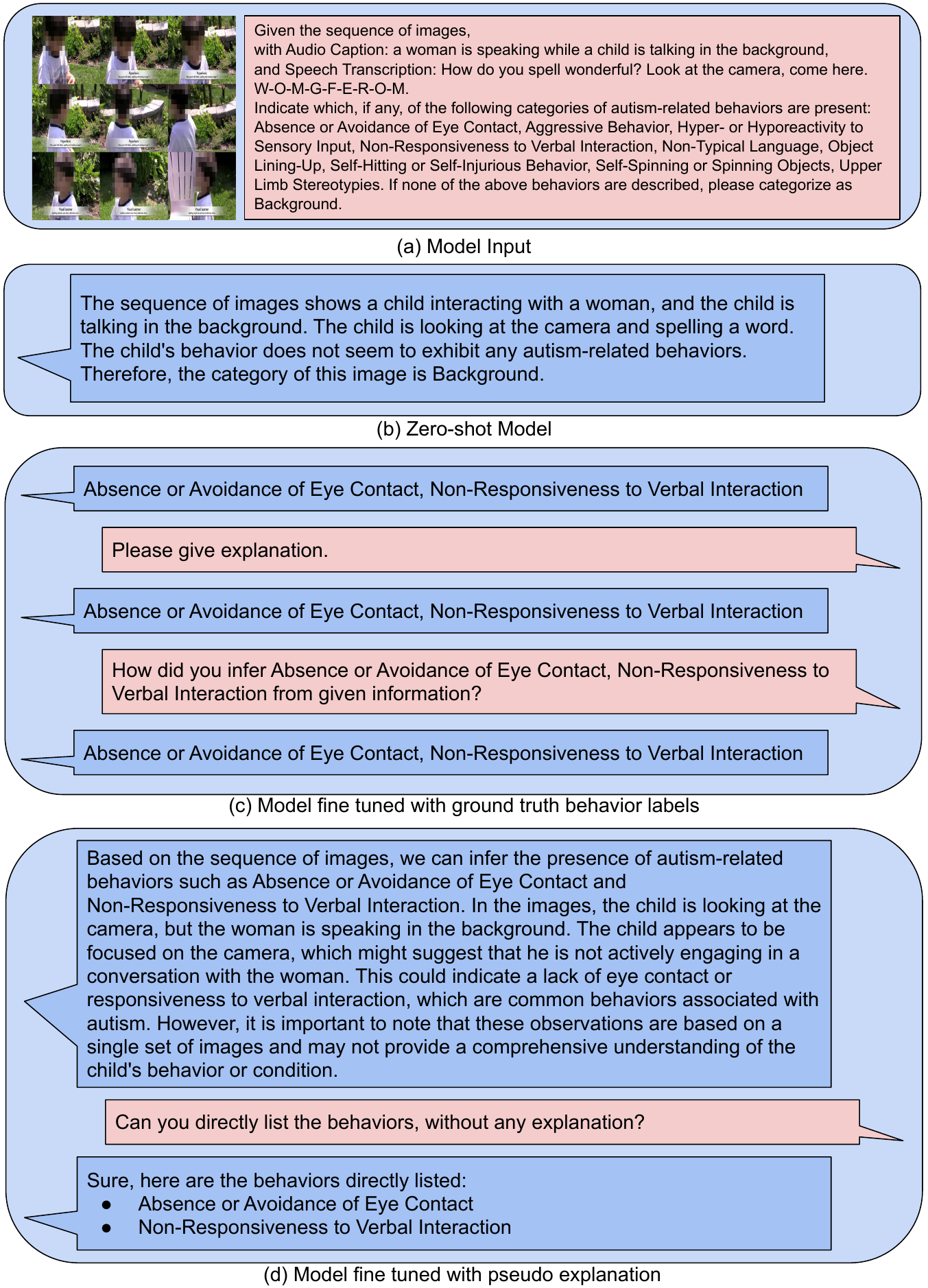}
  \vspace{-7mm}
  \caption{Explainability of different models. In this example, the ground truth labels are: \textit{Absence or Avoidance of Eye Contact} and \textit{Non-Responsiveness to Verbal Interaction}. We can see that the LLaVA model with zero-shot cannot even predict the correct behaviors; the one tuned directly on ground truth labels cannot explain the decision-making and has a catastrophic forgetting issue. Our model tuned with \textit{post-hoc to ad-hoc} framework correctly predicts the behaviors and gives explanations. It also follows other instructions well.}
  \label{fig:why_exp}
  \vspace{-5mm}
\end{figure}

\subsection{Beyond Recognition: Explanibility}
For an AI assistant to effectively aid doctors in autism screening, it is essential for the system to provide explanations for its suggestions. MLLMs demonstrate strong reasoning capabilities and can generate detailed explanations alongside predictions. However, zero-shot models may yield incorrect autism behavior recognition results, as illustrated in Figure~\ref{fig:why_exp}(b). While instruction tuning using ground truth behavior labels improves performance, this approach risks reducing the model to a mere classifier, potentially losing its reasoning ability due to catastrophic forgetting of previous knowledge (refer to Figure~\ref{fig:why_exp}(c)). To overcome this issue and achieve explainable predictions, we initiate an exploration, laying groundwork for further advancements in future research.

A straightforward solution to ensure accurate explanations is to utilize ground truth annotations from human experts for fine-tuning. However, this process is labor-intensive and costly. To circumvent these challenges, we propose a novel, efficient self-supervised pipeline: \textit{post-hoc to ad-hoc}, illustrated in Fig.~\ref{fig:exp}.
It contains two steps. First, with a visual input, audio caption, speech transcription, and prompt instruction $P_{\text{inst}}$, accompanied by ground truth behavior labels \( L_{gt} \), we employ LLaVA to infer explanations for these labels. The output of this step, termed \text{post-hoc} reasoning:
    \begin{equation}
        \text{R}_{\text{post-hoc}} = \text{MLLM}(I_V, AC, ST, P_{\text{inst}}, L_{gt})
    \end{equation}
    Second, in the absence of ground truth, we utilize the \text{post-hoc} reasoning as pseudo labels to train our model for generating ad-hoc reasoning, aiming for outputs similar to the \text{post-hoc} reasoning:
    \begin{equation}
        \text{R}_{\text{ad-hoc}} = \text{MLLM}(I_{V}, AC, ST, \hat{P}_{\text{inst}}; \theta)
    \end{equation}
    \begin{equation}
        \hat{\theta} = \arg\min_{\theta} CE(\text{R}_{\text{ad-hoc}}, \text{R}_{\text{post-hoc}})
    \end{equation}
    Here, \( \hat{P}_{\text{inst}} \) denotes the prompt for the ad-hoc step and $CE(\cdot)$ is cross-entropy loss.

Figure \ref{fig:why_exp}(d) illustrates the explanation results generated by our explainable framework. We can see that fine-tuning MLLMs with synthetic \text{post-hoc} reasoning data effectively prevents the model from reducing to a trivial behavior classifier. Additionally, it significantly improves its explainability for the task of recognizing autism behaviors.

\section{Future Directions}
Opportunities for expanding the size of the dataset include using an Activity Net~\cite{caba2015activitynet}-like system or active learning, similar to~\cite{kirillov2023segment}. Relatedly, conducting future keyword searches in languages other than English should help continue to diversify the dataset.
In our next project, we plan to leverage the many neurotypical videos available on social media platforms by approaching autism behavior detection through the lens of anomaly detection~\cite{gao2023unsupervised} or out-of-distribution problem-solving.

Finally, a crucial component of ASD screening is identifying the absence of certain social behaviors. This underlines the necessity of tools adept at discerning prolonged behavioral absences rather than immediate behavioral presences. To address this in the future, we hope to analyze longer, high-quality videos.

\section{Conclusion}
In this article, we present the AV-ASD dataset, a unique and comprehensive collection featuring social behavioral categories and repetitive behaviors. Our thorough experiments reveal that the integration of visual, audio, and speech data markedly improves autism behavior recognition, thereby facilitating the creation of more effective diagnostic tools. Our LLaVA-ASD model, which combines audio captioning and speech transcription with instruction tuning, excels in utilizing multimodal information for enhanced autism behavior recognition. Additionally, our \textit{post-hoc to ad-hoc} framework represents a pioneering attempt to tackle the challenge of explainability in autism behavior recognition.

\section*{Acknowledgment}

We express our deepest gratitude to Dr. Carolyn Garver of the Autism Treatment Center for her invaluable suggestions during the project, and to Cristina Rangel Uribe from the Social Communication Lab for her meticulous verification of all annotations to ensure their accuracy.

\ifCLASSOPTIONcaptionsoff
  \newpage
\fi

\bibliographystyle{IEEEtran}
\bibliography{main}

\begin{thebibliography}{10}
\providecommand{\url}[1]{#1}
\csname url@samestyle\endcsname
\providecommand{\newblock}{\relax}
\providecommand{\bibinfo}[2]{#2}
\providecommand{\BIBentrySTDinterwordspacing}{\spaceskip=0pt\relax}
\providecommand{\BIBentryALTinterwordstretchfactor}{4}
\providecommand{\BIBentryALTinterwordspacing}{\spaceskip=\fontdimen2\font plus
\BIBentryALTinterwordstretchfactor\fontdimen3\font minus \fontdimen4\font\relax}
\providecommand{\BIBforeignlanguage}[2]{{%
\expandafter\ifx\csname l@#1\endcsname\relax
\typeout{** WARNING: IEEEtran.bst: No hyphenation pattern has been}%
\typeout{** loaded for the language `#1'. Using the pattern for}%
\typeout{** the default language instead.}%
\else
\language=\csname l@#1\endcsname
\fi
#2}}
\providecommand{\BIBdecl}{\relax}
\BIBdecl

\bibitem{dsm5tr}
APA, \emph{Diagnostic and Statistical Manual of Mental Disorders}, 5th~ed.\hskip 1em plus 0.5em minus 0.4em\relax Washington, DC: American Psychiatric Association, 2022.

\bibitem{bottema2021avoiding}
K.~Bottema-Beutel, S.~K. Kapp, J.~N. Lester, N.~J. Sasson, and B.~N. Hand, ``Avoiding ableist language: Suggestions for autism researchers,'' \emph{Autism in adulthood}, vol.~3, no.~1, pp. 18--29, 2021.

\bibitem{fuller2020effects}
E.~A. Fuller and A.~P. Kaiser, ``The effects of early intervention on social communication outcomes for children with autism spectrum disorder: A meta-analysis,'' \emph{Journal of autism and developmental disorders}, vol.~50, pp. 1683--1700, 2020.

\bibitem{nahmias2019effectiveness}
A.~S. Nahmias, M.~Pellecchia, A.~C. Stahmer, and D.~S. Mandell, ``Effectiveness of community-based early intervention for children with autism spectrum disorder: A meta-analysis,'' \emph{Journal of Child Psychology and Psychiatry}, vol.~60, no.~11, pp. 1200--1209, 2019.

\bibitem{van2021age}
M.~van’t Hof, C.~Tisseur, I.~van Berckelear-Onnes, A.~van Nieuwenhuyzen, A.~M. Daniels, M.~Deen, H.~W. Hoek, and W.~A. Ester, ``Age at autism spectrum disorder diagnosis: A systematic review and meta-analysis from 2012 to 2019,'' \emph{Autism}, vol.~25, no.~4, pp. 862--873, 2021.

\bibitem{mchatrf2009}
D.~L. Robins, D.~Fein, and M.~Barton, ``M-chat-r/f: The modified checklist for autism in toddlers, revised with follow-up,'' Online, 2009, available at: https://www.mchatscreen.com/.

\bibitem{supekar2022deep}
K.~Supekar, C.~de~Los~Angeles, S.~Ryali, K.~Cao, T.~Ma, and V.~Menon, ``Deep learning identifies robust gender differences in functional brain organization and their dissociable links to clinical symptoms in autism,'' \emph{The British Journal of Psychiatry}, vol. 220, no.~4, pp. 202--209, 2022.

\bibitem{cilia2021computer}
F.~Cilia, R.~Carette, M.~Elbattah, G.~Dequen, J.-L. Gu{\'e}rin, J.~Bosche, L.~Vandromme, B.~Le~Driant \emph{et~al.}, ``Computer-aided screening of autism spectrum disorder: Eye-tracking study using data visualization and deep learning,'' \emph{JMIR human factors}, vol.~8, no.~4, p. e27706, 2021.

\bibitem{megerian2022evaluation}
J.~T. Megerian, S.~Dey, R.~D. Melmed, D.~L. Coury, M.~Lerner, C.~J. Nicholls, K.~Sohl, R.~Rouhbakhsh, A.~Narasimhan, J.~Romain \emph{et~al.}, ``Evaluation of an artificial intelligence-based medical device for diagnosis of autism spectrum disorder,'' \emph{NPJ digital medicine}, vol.~5, no.~1, p.~57, 2022.

\bibitem{jones2013attention}
W.~Jones and A.~Klin, ``Attention to eyes is present but in decline in 2--6-month-old infants later diagnosed with autism,'' \emph{Nature}, vol. 504, no. 7480, pp. 427--431, 2013.

\bibitem{mundy2022bidirectional}
P.~Mundy and J.~Bullen, ``The bidirectional social-cognitive mechanisms of the social-attention symptoms of autism,'' \emph{Frontiers in Psychiatry}, vol.~12, p. 752274, 2022.

\bibitem{openai2023gpt4}
OpenAI, ``Gpt-4 technical report,'' 2023.

\bibitem{li2023llava}
C.~Li, C.~Wong, S.~Zhang, N.~Usuyama, H.~Liu, J.~Yang, T.~Naumann, H.~Poon, and J.~Gao, ``Llava-med: Training a large language-and-vision assistant for biomedicine in one day,'' \emph{arXiv preprint arXiv:2306.00890}, 2023.

\bibitem{billing2020dream}
E.~Billing, T.~Belpaeme, H.~Cai, H.-L. Cao, A.~Ciocan, C.~Costescu, D.~David, R.~Homewood, D.~Hernandez~Garcia, P.~G{\'o}mez~Esteban \emph{et~al.}, ``The dream dataset: Supporting a data-driven study of autism spectrum disorder and robot enhanced therapy,'' \emph{PloS one}, vol.~15, no.~8, p. e0236939, 2020.

\bibitem{rehg2013decoding}
J.~Rehg, G.~Abowd, A.~Rozga, M.~Romero, M.~Clements, S.~Sclaroff, I.~Essa, O.~Ousley, Y.~Li, C.~Kim \emph{et~al.}, ``Decoding children's social behavior,'' in \emph{Proceedings of the IEEE conference on computer vision and pattern recognition}, 2013, pp. 3414--3421.

\bibitem{riva2020enigma}
G.~Riva, E.~Riva \emph{et~al.}, ``De-enigma: Multimodal human-robot interaction for teaching and expanding social imagination in autistic children,'' \emph{Cyberpsychology, behavior and social networking}, vol.~23, no.~11, pp. 806--807, 2020.

\bibitem{zunino2018video}
A.~Zunino, P.~Morerio, A.~Cavallo, C.~Ansuini, J.~Podda, F.~Battaglia, E.~Veneselli, C.~Becchio, and V.~Murino, ``Video gesture analysis for autism spectrum disorder detection,'' in \emph{2018 24th international conference on pattern recognition (ICPR)}.\hskip 1em plus 0.5em minus 0.4em\relax IEEE, 2018, pp. 3421--3426.

\bibitem{pandey2020guided}
P.~Pandey, A.~Prathosh, M.~Kohli, and J.~Pritchard, ``Guided weak supervision for action recognition with scarce data to assess skills of children with autism,'' in \emph{Proceedings of the AAAI Conference on Artificial Intelligence}, 2020, pp. 463--470.

\bibitem{del2017study}
M.~Del~Coco, M.~Leo, P.~Carcagn{\`\i}, F.~Fama, L.~Spadaro, L.~Ruta, G.~Pioggia, and C.~Distante, ``Study of mechanisms of social interaction stimulation in autism spectrum disorder by assisted humanoid robot,'' \emph{IEEE Transactions on Cognitive and Developmental Systems}, vol.~10, no.~4, pp. 993--1004, 2017.

\bibitem{dawson2018atypical}
G.~Dawson, K.~Campbell, J.~Hashemi, S.~J. Lippmann, V.~Smith, K.~Carpenter, H.~Egger, S.~Espinosa, S.~Vermeer, J.~Baker \emph{et~al.}, ``Atypical postural control can be detected via computer vision analysis in toddlers with autism spectrum disorder,'' \emph{Scientific reports}, vol.~8, no.~1, p. 17008, 2018.

\bibitem{martin2018objective}
K.~B. Martin, Z.~Hammal, G.~Ren, J.~F. Cohn, J.~Cassell, M.~Ogihara, J.~C. Britton, A.~Gutierrez, and D.~S. Messinger, ``Objective measurement of head movement differences in children with and without autism spectrum disorder,'' \emph{Molecular autism}, vol.~9, pp. 1--10, 2018.

\bibitem{li2023mmasd}
J.~Li, V.~Chheang, P.~Kullu, E.~Brignac, Z.~Guo, K.~E. Barner, A.~Bhat, and R.~L.~B. Name, ``Mmasd: A multimodal dataset for autism intervention analysis,'' \emph{arXiv preprint arXiv:2306.08243}, 2023.

\bibitem{rajagopalan2013self}
S.~Rajagopalan, A.~Dhall, and R.~Goecke, ``Self-stimulatory behaviours in the wild for autism diagnosis,'' in \emph{Proceedings of the IEEE International Conference on Computer Vision Workshops}, 2013, pp. 755--761.

\bibitem{negin2021vision}
F.~Negin, B.~Ozyer, S.~Agahian, S.~Kacdioglu, and G.~T. Ozyer, ``Vision-assisted recognition of stereotype behaviors for early diagnosis of autism spectrum disorders,'' \emph{Neurocomputing}, vol. 446, pp. 145--155, 2021.

\bibitem{wei2022vision}
P.~Wei, D.~Ahmedt-Aristizabal, H.~Gammulle, S.~Denman, and M.~A. Armin, ``Vision-based activity recognition in children with autism-related behaviors,'' \emph{arXiv preprint arXiv:2208.04206}, 2022.

\bibitem{ribeiro2023stimming}
G.~O. Ribeiro, M.~Grellert, and J.~T. Carvalho, ``Stimming behavior dataset-unifying stereotype behavior dataset in the wild,'' in \emph{2023 IEEE 36th International Symposium on Computer-Based Medical Systems (CBMS)}.\hskip 1em plus 0.5em minus 0.4em\relax IEEE, 2023, pp. 225--230.

\bibitem{deng2022problem}
A.~Deng, T.~Yang, C.~Chen, Q.~Chen, L.~Neely, and S.~Oyama, ``Problem behaviors recognition in videos using language-assisted deep learning model for children with autism,'' \emph{arXiv preprint arXiv:2211.09310}, 2022.

\bibitem{ali2022video}
A.~Ali, F.~F. Negin, F.~F. Bremond, and S.~Th{\"u}mmler, ``Video-based behavior understanding of children for objective diagnosis of autism,'' in \emph{VISAPP 2022-17th International Conference on Computer Vision Theory and Applications}, 2022.

\bibitem{rajagopalan2014detecting}
S.~S. Rajagopalan and R.~Goecke, ``Detecting self-stimulatory behaviours for autism diagnosis,'' in \emph{2014 IEEE International Conference on Image Processing (ICIP)}.\hskip 1em plus 0.5em minus 0.4em\relax IEEE, 2014, pp. 1470--1474.

\bibitem{kazakos2019epic}
E.~Kazakos, A.~Nagrani, A.~Zisserman, and D.~Damen, ``Epic-fusion: Audio-visual temporal binding for egocentric action recognition,'' in \emph{Proceedings of the IEEE/CVF International Conference on Computer Vision}, 2019, pp. 5492--5501.

\bibitem{xiao2020audiovisual}
F.~Xiao, Y.~J. Lee, K.~Grauman, J.~Malik, and C.~Feichtenhofer, ``Audiovisual slowfast networks for video recognition,'' \emph{arXiv preprint arXiv:2001.08740}, 2020.

\bibitem{gao2020listen}
R.~Gao, T.-H. Oh, K.~Grauman, and L.~Torresani, ``Listen to look: Action recognition by previewing audio,'' in \emph{Proceedings of the IEEE/CVF Conference on Computer Vision and Pattern Recognition}, 2020, pp. 10\,457--10\,467.

\bibitem{chen2022mm}
J.~Chen and C.~M. Ho, ``Mm-vit: Multi-modal video transformer for compressed video action recognition,'' in \emph{Proceedings of the IEEE/CVF Winter Conference on Applications of Computer Vision}, 2022, pp. 1910--1921.

\bibitem{lin2011error}
J.-C. Lin, C.-H. Wu, and W.-L. Wei, ``Error weighted semi-coupled hidden markov model for audio-visual emotion recognition,'' \emph{IEEE Transactions on Multimedia}, vol.~14, no.~1, pp. 142--156, 2011.

\bibitem{tao2020end}
F.~Tao and C.~Busso, ``End-to-end audiovisual speech recognition system with multitask learning,'' \emph{IEEE Transactions on Multimedia}, vol.~23, pp. 1--11, 2020.

\bibitem{tian2018audio}
Y.~Tian, J.~Shi, B.~Li, Z.~Duan, and C.~Xu, ``Audio-visual event localization in unconstrained videos,'' in \emph{Proceedings of the European Conference on Computer Vision (ECCV)}, 2018, pp. 247--263.

\bibitem{xue2021audio}
C.~Xue, X.~Zhong, M.~Cai, H.~Chen, and W.~Wang, ``Audio-visual event localization by learning spatial and semantic co-attention,'' \emph{IEEE Transactions on Multimedia}, vol.~25, pp. 418--429, 2021.

\bibitem{liu2022dense}
S.~Liu, W.~Quan, C.~Wang, Y.~Liu, B.~Liu, and D.-M. Yan, ``Dense modality interaction network for audio-visual event localization,'' \emph{IEEE Transactions on Multimedia}, 2022.

\bibitem{huang2023egocentric}
C.~Huang, Y.~Tian, A.~Kumar, and C.~Xu, ``Egocentric audio-visual object localization,'' in \emph{Proceedings of the IEEE/CVF Conference on Computer Vision and Pattern Recognition}, 2023, pp. 22\,910--22\,921.

\bibitem{mo2023audio}
S.~Mo and Y.~Tian, ``Audio-visual grouping network for sound localization from mixtures,'' in \emph{Proceedings of the IEEE/CVF Conference on Computer Vision and Pattern Recognition}, 2023, pp. 10\,565--10\,574.

\bibitem{jiang2023leveraging}
Y.~Jiang, J.~Yin, and Y.~Dang, ``Leveraging the video-level semantic consistency of event for audio-visual event localization,'' \emph{IEEE Transactions on Multimedia}, 2023.

\bibitem{tian2020unified}
Y.~Tian, D.~Li, and C.~Xu, ``Unified multisensory perception: Weakly-supervised audio-visual video parsing,'' in \emph{Computer Vision--ECCV 2020: 16th European Conference, Glasgow, UK, August 23--28, 2020, Proceedings, Part III 16}.\hskip 1em plus 0.5em minus 0.4em\relax Springer, 2020, pp. 436--454.

\bibitem{mo2022multi}
S.~Mo and Y.~Tian, ``Multi-modal grouping network for weakly-supervised audio-visual video parsing,'' in \emph{Advances in Neural Information Processing Systems}, 2022.

\bibitem{zhou2022audio}
J.~Zhou, J.~Wang, J.~Zhang, W.~Sun, J.~Zhang, S.~Birchfield, D.~Guo, L.~Kong, M.~Wang, and Y.~Zhong, ``Audio--visual segmentation,'' in \emph{European Conference on Computer Vision}.\hskip 1em plus 0.5em minus 0.4em\relax Springer, 2022, pp. 386--403.

\bibitem{zhao2018sound}
H.~Zhao, C.~Gan, A.~Rouditchenko, C.~Vondrick, J.~McDermott, and A.~Torralba, ``The sound of pixels,'' in \emph{Proceedings of the European conference on computer vision (ECCV)}, 2018, pp. 570--586.

\bibitem{gao2021visualvoice}
R.~Gao and K.~Grauman, ``Visualvoice: Audio-visual speech separation with cross-modal consistency,'' in \emph{2021 IEEE/CVF Conference on Computer Vision and Pattern Recognition (CVPR)}.\hskip 1em plus 0.5em minus 0.4em\relax IEEE, 2021, pp. 15\,490--15\,500.

\bibitem{su2023separating}
Y.~Su, A.~Vosoughi, S.~Deng, Y.~Tian, and C.~Xu, ``Separating invisible sounds toward universal audiovisual scene-aware sound separation,'' \emph{arXiv preprint arXiv:2310.11713}, 2023.

\bibitem{li2022learning}
G.~Li, Y.~Wei, Y.~Tian, C.~Xu, J.-R. Wen, and D.~Hu, ``Learning to answer questions in dynamic audio-visual scenarios,'' in \emph{Proceedings of the IEEE/CVF Conference on Computer Vision and Pattern Recognition}, 2022, pp. 19\,108--19\,118.

\bibitem{zhu2020describing}
Y.~Zhu, Y.~Wu, Y.~Yang, and Y.~Yan, ``Describing unseen videos via multi-modal cooperative dialog agents,'' in \emph{Computer Vision--ECCV 2020: 16th European Conference, Glasgow, UK, August 23--28, 2020, Proceedings, Part XXIII 16}.\hskip 1em plus 0.5em minus 0.4em\relax Springer, 2020, pp. 153--169.

\bibitem{alamri2019audio}
H.~Alamri, V.~Cartillier, A.~Das, J.~Wang, A.~Cherian, I.~Essa, D.~Batra, T.~K. Marks, C.~Hori, P.~Anderson \emph{et~al.}, ``Audio visual scene-aware dialog,'' in \emph{Proceedings of the IEEE/CVF Conference on Computer Vision and Pattern Recognition}, 2019, pp. 7558--7567.

\bibitem{radford2021learning}
A.~Radford, J.~W. Kim, C.~Hallacy, A.~Ramesh, G.~Goh, S.~Agarwal, G.~Sastry, A.~Askell, P.~Mishkin, J.~Clark \emph{et~al.}, ``Learning transferable visual models from natural language supervision,'' in \emph{International conference on machine learning}.\hskip 1em plus 0.5em minus 0.4em\relax PMLR, 2021, pp. 8748--8763.

\bibitem{girdhar2023imagebind}
R.~Girdhar, A.~El-Nouby, Z.~Liu, M.~Singh, K.~V. Alwala, A.~Joulin, and I.~Misra, ``Imagebind: One embedding space to bind them all,'' in \emph{Proceedings of the IEEE/CVF Conference on Computer Vision and Pattern Recognition}, 2023, pp. 15\,180--15\,190.

\bibitem{radford2023robust}
A.~Radford, J.~W. Kim, T.~Xu, G.~Brockman, C.~McLeavey, and I.~Sutskever, ``Robust speech recognition via large-scale weak supervision,'' in \emph{International Conference on Machine Learning}.\hskip 1em plus 0.5em minus 0.4em\relax PMLR, 2023, pp. 28\,492--28\,518.

\bibitem{liu2023llava}
H.~Liu, C.~Li, Q.~Wu, and Y.~J. Lee, ``Visual instruction tuning,'' 2023.

\bibitem{liu2023improvedllava}
H.~Liu, C.~Li, Y.~Li, and Y.~J. Lee, ``Improved baselines with visual instruction tuning,'' 2023.

\bibitem{labbe2023conette}
E.~Labb{\'e}, T.~Pellegrini, and J.~Pinquier, ``Conette: An efficient audio captioning system leveraging multiple datasets with task embedding,'' \emph{arXiv preprint arXiv:2309.00454}, 2023.

\bibitem{ouyang2022training}
L.~Ouyang, J.~Wu, X.~Jiang, D.~Almeida, C.~Wainwright, P.~Mishkin, C.~Zhang, S.~Agarwal, K.~Slama, A.~Ray \emph{et~al.}, ``Training language models to follow instructions with human feedback,'' \emph{Advances in Neural Information Processing Systems}, vol.~35, pp. 27\,730--27\,744, 2022.

\bibitem{dai2023instructblip}
W.~Dai, J.~Li, D.~Li, A.~M.~H. Tiong, J.~Zhao, W.~Wang, B.~Li, P.~Fung, and S.~Hoi, ``Instructblip: Towards general-purpose vision-language models with instruction tuning,'' 2023.

\bibitem{zhang2023video}
H.~Zhang, X.~Li, and L.~Bing, ``Video-llama: An instruction-tuned audio-visual language model for video understanding,'' \emph{arXiv preprint arXiv:2306.02858}, 2023.

\bibitem{liu2023visual}
H.~Liu, C.~Li, Q.~Wu, and Y.~J. Lee, ``Visual instruction tuning,'' \emph{arXiv preprint arXiv:2304.08485}, 2023.

\bibitem{hu2021lora}
E.~J. Hu, Y.~Shen, P.~Wallis, Z.~Allen-Zhu, Y.~Li, S.~Wang, L.~Wang, and W.~Chen, ``Lora: Low-rank adaptation of large language models,'' \emph{arXiv preprint arXiv:2106.09685}, 2021.

\bibitem{drossos2020clotho}
K.~Drossos, S.~Lipping, and T.~Virtanen, ``Clotho: An audio captioning dataset,'' in \emph{ICASSP 2020-2020 IEEE International Conference on Acoustics, Speech and Signal Processing (ICASSP)}.\hskip 1em plus 0.5em minus 0.4em\relax IEEE, 2020, pp. 736--740.

\bibitem{caba2015activitynet}
B.~G. Fabian Caba~Heilbron, Victor~Escorcia and J.~C. Niebles, ``Activitynet: A large-scale video benchmark for human activity understanding,'' in \emph{Proceedings of the IEEE Conference on Computer Vision and Pattern Recognition}, 2015, pp. 961--970.

\bibitem{kirillov2023segment}
A.~Kirillov, E.~Mintun, N.~Ravi, H.~Mao, C.~Rolland, L.~Gustafson, T.~Xiao, S.~Whitehead, A.~C. Berg, W.-Y. Lo \emph{et~al.}, ``Segment anything,'' \emph{arXiv preprint arXiv:2304.02643}, 2023.

\bibitem{gao2023unsupervised}
J.~Gao, X.~Jiang, Y.~Yang, D.~Li, and L.~Qiu, ``Unsupervised video anomaly detection for stereotypical behaviours in autism,'' in \emph{ICASSP 2023-2023 IEEE International Conference on Acoustics, Speech and Signal Processing (ICASSP)}.\hskip 1em plus 0.5em minus 0.4em\relax IEEE, 2023, pp. 1--5.

\end{thebibliography}

\end{document}


\title{Hear Me, See Me, Understand Me:\\ Audio-Visual Autism Behavior Recognition\\ (Supplementary Material)}

\author{
        Shijian~Deng,~\IEEEmembership{Member,~IEEE,}
        Erin E. Kosloski,
        Siddhi~Patel,
        Zeke A. Barnett,
        Yiyang~Nan, \\
        Alexander~Kaplan,
        Sisira~Aarukapalli,
        William T. Doan,
        Matthew~Wang, 
        Harsh~Singh, \\
        Pamela R. Rollins, Yapeng~Tian,~\IEEEmembership{Member,~IEEE} 

\thanks{S. Deng, Z. Barnett, A. Kaplan, S. Aarukapalli, W Doan, M. Wang, and Y. Tian are with the Department of Computer Science, The University of Texas at Dallas, Richardson, TX, 75080 USA.
}
\thanks{E. Kosloski and P. Rollins are with the School of Behavioral and Brain Sciences, The University of Texas at Dallas, Dallas, TX, 75235 USA 
.}
\thanks{Y. Nan is with the Department of Computer Science, Brown University, Providence, RI, 02912 USA 
.}
\thanks{H. Singh is with the Mohamed bin Zayed University of Artificial Intelligence, Abu Dhabi, UAE
.}
}

\maketitle

\section{Behavior Descriptions}
\label{sec:behavior_descriptions}

For each category of behavior in the AV-ASD dataset, we have a detailed description that is written by a qualified expert in \cref{tab:descriptions}.

\begin{table*}
  \centering
  \small
  \begin{tabular}{@{}p{5cm}p{11cm}@{}}
    \toprule
    Behavior & Description \\
    \midrule
    Absence or Avoidance of Eye Contact & Difficulty with or avoidance of eye contact is a common behavior in autistic individuals. They may not naturally use eye contact as a social cue in the same way that neurotypical individuals do, and some may find maintaining eye contact uncomfortable or even distressing.\\\midrule
    Non-Responsiveness to Verbal Interaction & Many autistic people have challenges with verbal communication. This may take the form of not responding when spoken to, not initiating verbal interactions, or not using verbal communication in a typical way. This can be due to a variety of reasons, including difficulty producing language, problems with auditory processing, or discomfort with social interaction. Some autistic people may be consistently minimally verbal or nonspeaking, while others may become temporarily unable to speak in stressful situations.\\\midrule
    Non-Typical Language & Some autistic individuals use language in unconventional ways, which can include echolalia (repeating words or phrases), limited expressive vocabulary, idiosyncratic language (unique phrases or words), and literal interpretation of speech. These instances of non-typical language use can occur for a variety of reasons.\\\thickmidrule 
    Aggressive Behavior & Some autistic individuals may exhibit aggressive behavior, which can manifest as physical aggression (\textit{e.g.,} hitting, kicking, biting), verbal aggression (\textit{e.g.,} shouting, insults), or destructive behavior (\textit{e.g.,} breaking objects). Such aggression may be a response to a range of factors, such as overwhelming sensory environments, difficulty with social interactions, or frustration stemming from communication difficulties.\\\midrule
    Self-Hitting or Self-Injurious Behavior & Some individuals on the autism spectrum may engage in self-injurious behaviors, such as head banging or hitting themselves. These behaviors can serve a number of functions, including self-stimulation, a reaction to sensory overload, a means of expressing distress, or a way to communicate needs.\\\midrule
    Hyper- or Hyporeactivity to Sensory Input & Autistic individuals often have non-typical responses to sensory input. Hyper-reactivity refers to an oversensitivity to sensory stimuli, such as finding lights too bright or noises too loud. Hyporeactivity, on the other hand, refers to an undersensitivity, where an individual may not respond to certain stimuli or may seek out intense sensory experiences.\\\midrule
    Object Lining-Up & Some autistic individuals may display a behavior known as lining up or ordering objects. This could manifest as arranging toys in a precise pattern or lining up items in a straight line. This behavior may be comforting or soothing, provide a sense of order and predictability, or satisfy a desire for visual or spatial symmetry.\\\midrule
    Self-Spinning or Spinning Objects & Spinning or twirling, either in place or with objects, is another behavior sometimes observed in individuals on the autism spectrum. This could be a form of self-stimulation (also known as stimming), providing sensory input that is calming or focusing, or this behavior could be a way of coping with a stressful or overstimulating environment.\\\midrule
    Upper Limb Stereotypies & Stereotypies are repetitive, rhythmic movements that can often be observed in autistic individuals. Upper limb stereotypies can involve flapping hands, twisting wrists, or other repetitive movements of the arms or hands. These movements are generally thought to serve a self-soothing or self-stimulating function, or they may be a response to excitement or stress.\\\midrule
    Background & None of the above behaviors are present in the clip.\\
    \bottomrule
  \end{tabular}
  \caption{Description of each behavior category. Note that the first 3 categories represent social behaviors.}
  \label{tab:descriptions}
\end{table*}

\section{Keyword Searches}
\label{sec:keywords}

To gather clips of autistic behaviors for the dataset, we conducted searches on Youtube and FaceBook using the following keywords: 'autism', 'autistic', 'asperger', 'asd', 'adhd', 'hand flapping autism', 'tiptoes autism', 'head banging autism', 'fussiness autism', 'screaming autism', 'pouring drinks back and forth autism', 'biting autism', 'aggressiveness autism', 'lacks response to voice commands autism', 'lacks response to verbal commands autism', 'avoids eye contact autism', 'lack of speech autism',  'lack of communication autism', 'problems with food autism', 'problems with textures autism', 'problems with clothing autism', 'child head banging', 'child spinning'

To gather clips for the "Background" category of the dataset, we used the following keywords: 'child', 'children', 'kid', 'kids', 'toddler', 'toddlers', 'baby', and 'babies.'

\ifCLASSOPTIONcaptionsoff
  \newpage
\fi